\documentclass[12pt]{article}
\input epsf.sty
\def\ZZZ{{\hbox{ Z\kern-1.6mm Z}}}

\newcommand{\tl}{\wt\lambda}

\newcommand{\wt}{\widetilde}

\newcommand{\TT}{{\cal T}}

\newcommand{\be}{\begin{equation}}
\newcommand{\ee}{\end{equation}}
\newcommand{\ben}{\begin{eqnarray}\displaystyle}
\newcommand{\een}{\end{eqnarray}}
\newcommand{\refb}[1]{(\ref{#1})}
\newcommand{\p}{\partial}

\def\one{{\hbox{ 1\kern-.8mm l}}}
\def\zero{{\hbox{ 0\kern-1.5mm 0}}}
 
\begin{document}
{}~
{}~
\hfill\vbox{\hbox{hep-th/0312153}
}\break
 
\vskip .6cm
\centerline{\Large \bf 
Remarks on Tachyon Driven Cosmology}

\vskip .6cm
\medskip

\vspace*{4.0ex}
 
\centerline{\large \rm
Ashoke Sen}
 
\vspace*{4.0ex}

\centerline{\large \it Harish-Chandra Research Institute}

\centerline{\large \it  Chhatnag Road, Jhusi,
Allahabad 211019, INDIA}
 
\centerline{E-mail: ashoke.sen@cern.ch,
sen@mri.ernet.in}
 
\vspace*{5.0ex}
 
\centerline{\bf Abstract} \bigskip

We begin by reviewing the results on the decay of unstable D-branes in
type II string theory, and the open-closed string duality proposal that
arises from these studies. We then apply this proposal to the study of
tachyon driven cosmology, namely cosmological solutions describing the
decay of unstable space filling D-branes. This naturally gives rise to a
time reversal invariant bounce solution with positive spatial curvature.
In the absence of a bulk cosmological constant the universe always begins
with a big bang and ends in a big crunch. In the presence of a bulk
cosmological constant one may get non-singular cosmological solutions for
some special range of initial conditions on the tachyon.

\vfill \eject
 
\baselineskip=16.2pt


\noindent{\bf Introduction and review of classical dynamics of open 
string 
tachyon}

\medskip

Type IIA and IIB string theories contain in their spectrum unstable 
non-BPS D-branes besides the stable BPS D-branes. These unstable branes 
are characterized by having a single tachyonic mode of mass$^2=-{1\over 
2}$ (in $\alpha'=1$ unit) living on their world-volume. The tachyon 
potential $V(T)$, which is even under the change of sign of the tachyon 
field 
$T$, has a local maximum at $T=0$ and a pair of global minima at $\pm 
T_{min}$ where the negative contribution from 
the potential exactly cancels the tension of the D-brane. As a result 
the configuration where the tachyon potential is at its minimum 
corresponds to vacuum without any D-branes, and fluctuations of the open 
string field around these minima do not contain any perturbative open 
string states in the spectrum\cite{initial}.

One can study time dependent solutions where the tachyon rolls away from 
near the top of the potential towards the minimum of the potential. If we 
consider such a rolling tachyon solution on an unstable D$p$-brane located 
at $x^\alpha=0$ ($(p+1)\le\alpha\le 9$), where we denote by 
$x^\alpha$ the coordinates transverse to the D-brane, by $x^0$ the  time 
coordinate and by $x^i$ ($1\le i\le 
p$) the spatial directions tangential to the D$p$-brane, then the energy 
momentum 
tensor associated with the rolling tachyon solution takes the 
form\cite{0203211,0203265}:
\be \label{ess1}
T_{00}=\hbox{constant}, \quad T_{ij}(x^0) = p(x^0) \, \delta_{ij} \,
\delta(\vec x_\perp), \quad
T_{i0}=T_{0\alpha} = T_{i\alpha} =
T_{\alpha\beta}=0\, .
\ee
Here $p(x^0)$ is a function of the time coordinate $x^0$ and is commonly 
known as the pressure of the system. $\vec x_\perp=(x^{p+1}, \ldots 
x^9)$ denotes the coordinate vector transverse to the D$p$-brane. Another 
important 
quantity is the 
dilaton charge density associated with the rolling tachyon configuration, 
defined as the coupling of the rolling tachyon configuration to the 
dilaton field at fixed {\it string metric}. This takes the form:
\be \label{ess2}
Q(x^0) \, \delta(\vec x_\perp)\, .
\ee
Explicit forms of $p(x^0)$ and $Q(x^0)$ for a given $T_{00}$ is 
known\cite{0203211,0203265,electric,rolling}. To 
specify them we need to consider two cases separately. 
\begin{enumerate}
\item First consider the case where the total energy density $T_{00}$ of 
the system is less than the 
tension $\wt\TT_p$ of the non-BPS D$p$-brane. In this case we can 
parametrize $T_{00}$ as 
\be \label{ess3}
T_{00} = \wt\TT_p \, \cos^2(\pi\tl)\, .
\ee
$p(x^0)$ and $Q(x^0)$ for this system are given by\cite{0203265}
\ben \label{ess4}
&& p(x^0) = - {\wt\TT_p} \, f(x^0), \qquad Q(x^0) = \wt\TT_p f(x^0)\, , 
\nonumber \\
&& f(x^0) = {1\over 1 + e^{\sqrt 2 x^0} \sin^2(\tl\pi)} + {1
\over
1 + e^{-\sqrt 2 x^0} \sin^2(\tl\pi)} - 1\, .
\een
For any fixed $\tl$, $f(x^0)\to 0$ as $x^0\to\infty$, and hence the 
pressure and the dilaton charge density vanishes asymptotically. The other 
imporatnt point to note is that at $\tl={1\over 2}$ all components of the 
stress tensor and the dilaton charge density vanish, indicating that 
this gives the vacuum without any D-brane.

\item Next we consider the case where the total energy density $T_{00}$ of
the system is larger than the
tension $\wt\TT_p$ of the non-BPS D$p$-brane. In this case we can
parametrize $T_{00}$ as
\be \label{ess5}
T_{00} = \wt\TT_p \, \cosh^2(\pi\tl)\, .
\ee
$p(x^0)$ and $Q(x^0)$ for this system are given by\cite{0203265}
\ben \label{ess6}
&& p(x^0) = - {\wt\TT_p} \, f(x^0), \qquad Q(x^0) = \wt\TT_p f(x^0)\, , 
\nonumber \\
&& f(x^0) = {1\over 1 + e^{\sqrt 2 x^0} \sinh^2(\tl\pi)} + {1
\over
1 + e^{-\sqrt 2 x^0} \sinh^2(\tl\pi)} - 1\, .
\een
Again we see that for any fixed $\tl$, $f(x^0)\to 0$ as $x^0\to\infty$, 
and hence the
pressure and the dilaton charge density vanish asymptotically.

\end{enumerate}

\medskip

\noindent{\bf Coupling to closed strings and open-closed string duality 
conjecture}

\medskip

So far in our discussion we have ignored the coupling of the D-brane to 
closed strings. Since the rolling tachyon background acts as a time 
dependent source for various closed string fields, we expect that there 
will be closed string radiation from the D-brane as the tachyon rolls down 
towards the minimum of the potential. However, in the weak coupling limit 
we would expect this effect to be small, since the closed strings 
couple to the 
D-brane at order $g$ where $g$ is the string coupling constant. Explicit 
calculation\cite{0209222,0303139,0304192} 
shows 
that this is 
indeed the case for individual closed string modes; the fraction of the 
D-brane energy carried away by each closed string mode is of order $g$ 
and falls off exponentially as we go to higher massive closed string 
states. But due to the rapid growth of closed string density of states at 
large mass level, the picture changes dramatically when we sum over all 
the closed string modes.
The results of \cite{0303139,0304192} for emission of closed strings from 
unstable D-branes all of whose spatial world-volume directions lie along a 
compact torus can be 
summarized as follows:
\begin{enumerate}
{\item All the energy of the D-brane is radiated away into closed
strings.}

{\item Most of the emitted energy is carried by closed strings of
mass $\sim {1 / g}$.}

{\item Typical momentum transverse to the brane, and typical 
winding charge along the brane of an emitted closed string is of order 
$1/\sqrt{g}$. This in particular shows that the typical velocity of a 
closed string along directions transverse to the brane is of order 
$\sqrt{g}$.}

\end{enumerate}

Given that all the energy of the D-brane is carried away by the closed 
strings, it would seem that the effect of closed string emission 
invalidates the open string analysis. However, before we 
make such a 
conclusion, let us compare the  the properties of the emitted closed 
strings
with
those infered from the tree level open string analysis\cite{0305011}. 
First of all, tree 
level open string analysis tells us that the final system
has:
\be \label{est1}
Q / T_{00} = 0\, .
\ee
On the other hand by examining the closed string world-sheet action in the 
background  {\it string
metric} $G_{\mu\nu}$, the anti-symmetric tensor field $B_{\mu\nu}$ and the
dilaton $\Phi$ at zero momentum,
\be \label{est2}
S_{world-sheet} = \int d^2 z (G_{\mu\nu}(X) + B_{\mu\nu}(X))
\p_z X^\mu
\p_{\bar z} X^\nu\, ,
\ee
we see that the closed string world-sheet does not
couple to the zero momentum dilaton, provided we take the {\it string 
metric}, the anti-symmetric tensor field and the 
dilaton as independent 
fields.
This shows the final state closed strings carry zero total dilaton
charge. Hence the dilaton charge of the final state closed
strings
agrees with that computed in the open string description.

Next we note that the tree level open string analysis tells us that the 
final system
has:
\be \label{est3}
p / T_{00} = 0\, .
\ee
On the other hand, closed string analysis tells us that the final
closed
strings have mass $m$ of order $1/g$, momentum $k_\perp$
transverse to the D-brane of order $1/\sqrt{g}$ and 
winding $w_\parallel$ tangential to the D-brane of order $1/\sqrt{g}$. 
For such a system the ratio of transverse pressure to the energy density 
is of order $(k_\perp/m)^2\sim g$ and the ratio of tangential pressure to 
the energy density is of order $-(w_\parallel/m)^2\sim -g$.
Since both these ratios 
vanish in the $g\to 0$ limit, we again see that the pressure of the 
final state closed strings matches the
result
computed in the open string description.

Such agreements between open and closed string results also hold
for
more general cases, {\it e.g.} in the decay of unstable branes in the
presence of electric field. Consider, for example, the decay of a 
D$p$-brane along $x^1,\ldots x^p$ plane, with
an
electric field $e$ along the $x^1$ axis. In this case the final state is 
characterized by its energy-momentum tensor
$T_{\mu\nu}$, source $S_{\mu\nu}$ for anti-symmetric tensor field 
$B_{\mu\nu}$ and the
dilaton charge $Q$. One can show that in the $x^0\to\infty$ 
limit\cite{electric}:
\be \label{est4}
T^{00}=|\Pi|\, e^{-1} \, , \quad
T^{11} = - |\Pi| \, e\, , \quad
S^{01} = \Pi \, ,
\ee
where $\Pi$ is a parameter labelling the solution. All other components of 
$T_{\mu\nu}$ and $S_{\mu\nu}$, as well as the dilaton charge vanishes in 
this limit. It can be shown that these tree level open string results 
again agree exactly with 
the
properties of the final state closed strings into which the D-brane
decays\cite{0306137}.

Since in all these cases the tree level open string results for various 
properties of the final state agree with the properties of the closed 
strings produced in the decay of the brane, we are led to conjecture that 
the tree level open string theory provides a
description of the rolling tachyon system which is {\it dual} to the
description in terms of closed  string 
emission\cite{0305011,0306137}.\footnote{This 
correspondence has been checked only for the space averaged values of 
various quantities, and not for example, for the local distribution of the 
various charges like the stress tensor, dilaton charge and anti-symmetric 
tensor field charge.  This is due to the fact that we can easily give a 
gauge invariant definition of the space-averaged quantities since they 
are measured by coupling to on-shell (zero momentum) closed string states, 
but it 
is more difficult to give a gauge invariant definition of local 
distribution of these charges\cite{0308068}.}$^,$\footnote{The precise 
mechanism of how the closed strings emerge from open string theory is 
still not understood. For some ideas see \cite{duality,0305011}.}
This is different from the usual duality between closed strings and open 
strings on stable D-branes 
where 
{\it one loop} open string theory contains information about closed 
strings. The difference is likely to be due to the fact that in the latter 
case the perturbative excitations around the stable minimum of the 
potential are open string excitations, whereas in the former case 
excitations around the stable minimum of the
potential are closed string excitations.

In 
order to put the tree level open-closed string duality conjecture on a 
firmer footing, one must show that it 
arises from a more complete conjecture
involving full quantum open string theory on an unstable D-brane. The full 
conjecture, suggested in \cite{0308068}, takes the following form:

{\it There is a quantum open string field theory
(OSFT) that describes the full dynamics of an unstable Dp-brane without an
explicit coupling to closed strings. Furthermore, Ehrenfest theorem holds
in the weakly coupled OSFT; the classical results correctly
describe the evolution of the quantum expectation values.}

Note that this conjecture {\it does not} imply that the quantum open 
string theory on a given system of unstable
D-branes gives a complete description of the {\it full string theory}. It 
only 
states that this open string field theory describes a quantum mechanically 
consistent subsector of the full string theory, and
is fully capable of describing the quantum decay process of an unstable 
D-brane.
One of the consequences of
this correspondence is that the notion of naturalness of a solution may
differ dramatically in the open and the closed string description. A
solution describing the decay of a single (or a few) unstable D-branes may
look highly contrived in the closed string description, since 
a
generic deformation of this background in closed string theory may not be 
describable by the dynamics of a single D-brane, and may require
a large (or even infinite) number of D-branes for its description in open
string theory.

\medskip

\noindent{\bf Lessons from $c=1$ string theory}

\medskip

A concrete illustration of the open-closed string duality may be found in 
the $c=1$ string 
theory\cite{0305159,c=1,0308068}. The states in this theory are in one to 
one 
correspondence
with the states of the free fermion field theory, with each fermion moving
in an inverted harmonic oscillator potential:
\be \label{e1}
V(q) = - {1\over 2} q^2 + {1\over g} \, .
\ee
The ground state of the theory corresponds to all levels below zero 
being filled 
and all levels above zero being empty. 
The closed string states of the theory are related to the bosons obtained
by bosonizing the fermion field\cite{bosonize}. On the other hand,
the states of a single D0-brane of this theory
correspond to a single fermion 
excited from the fermi level to a level above zero. 
Thus the quantum dynamics of a single D0-brane is described by an inverted 
harmonic oscillator Hamiltonian with potential given in \refb{e1}, with 
a sharp 
cut-off on the energy 
eigenvalue ($E\ge 0$) due to the Pauli exclusion principle. 
This is a fully consistent quantum system. However,
this 
set of states clearly
form only a small subset of all the quantum excitations of the full $c=1$ 
string theory. In 
particular we
cannot describe generic closed string excitations involving particle-hole 
pair creation as a state of this single D0-brane quantum 
mechanics\cite{0308068}.
This makes the description of a D0-brane state in closed string theory 
look highly contrived, since this is a highly specialized coherent state 
of 
closed strings in which a finely tuned coherent closed string radiation 
form the D0-brane and decays back into a time reversed version of the 
initial incoming radiation\cite{0303139,0304192}.\footnote{Such classical 
solutions have been called S-brane in the literature\cite{sbrane}.} Any 
generic deformation of this 
incoming 
closed 
string radiation will give rise to a state that is not describable as 
single fermion excitations of the $c=1$ string theory, {\it i.e.} as a 
state of 
a single D0-brane.

The lesson to be taken back from here is that if we are trying to find a 
closed string description of a D-brane decay process, we may have to forgo 
our 
usual notion of naturalness, and look for highly specialized closed string 
configurations. This does not mean that a generic closed string 
deformation of such a state is not an allowed state of string theory; but 
it is not an allowed state of the open string theory 
describing the D-brane under consideration. We shall now use this insight 
for describing the gravitational field 
configuration associated with the decay of a space-filling D-brane.

\medskip

\noindent{\bf Tachyon driven cosmology}

\medskip

The system that we have in mind is a superstring theory compactified on a 
six dimensional compact manifold so that we have 3 large spatial 
directions. We shall assume that all the moduli are frozen, and that the 
universe has a bulk cosmological constant $\Lambda$, as in the KKLT 
model\cite{0301240}. In this theory we take an unstable D-brane that 
extends along the three large spatial directions. If for example we are 
considering type IIB string theory as in \cite{0301240}, we can take an 
unstable D4-brane wrapped on a one cycle that arises by modding out the 
Calabi-Yau space by a discrete subgroup.\footnote{We shall not worry about 
the relative magnitude of various mass scales and try to justify the use 
of the action \refb{e7} from first principles; instead we shall simply use 
\refb{e7} as our starting point as has been done in all the recent 
papers on tachyon matter cosmology.} In this case, following 
\cite{effective,0303139,0204008,tachcosmo,0205121,0303035} we shall model 
this 
system by 
the 
effective action:
\ben \label{e7}
S &=& - {1\over 16 \pi G} \, \int d^4 x \, \bigg[- \sqrt{-\det g} \, R + 
 V(T) \sqrt{-\det(g_{\mu\nu}+ \p_\mu T
\p_\nu
T)} + \Lambda\sqrt{-\det g} \bigg]\, , \nonumber \\ \cr
&& V(T) = V_0 / \cosh(T /\sqrt 
2)\, ,
\een 
where $V_0\sim g^{-1}$ is the mass per unit three volume of the 
unstable brane and the Newton's constant $G$ is of order $g^2$. 
We shall assume that $\Lambda << V_0$.
We look for a spatially homogeneous time dependent solution of this 
equation in the Friedman-Robertson-Walker (FRW) form:
\ben \label{e3}
T &=& T(x^0), \nonumber \\
ds^2 &=& - (dx^0)^2 + a(x^0)^2 \left( {dr^2\over 
1-kr^2} 
+ r^2 d\theta^2 + \sin^2\theta d\phi^2\right), \quad k=0 
\quad \hbox{or} \quad
\pm
1\, .
\een
The equations of motion for $a$ and $T$ derived from the action \refb{e7} 
take the form:
\ben \label{e8}
{\ddot a\over a} &=& {8 \pi G\over 3} \, \bigg[ \Lambda + {V(T)\over
\sqrt{1 -
\dot T^2}} - {3\over 2} \, {\dot T^2 V(T) \over \sqrt{1 -
\dot T^2}}\bigg] \nonumber \\
\ddot T &=& - (1 - \dot T^2) \bigg[ {V'(T)\over V(T)} + 3 \, \dot
T \,
{\dot a
\over
a}\bigg]\, .
\een
The Friedman equation, obtained by varying the action with respect to 
$g_{00}$, takes the form:
\be \label{e9}
\left({\dot a \over a}\right)^2 = - {k\over a^2} + {8\pi G\over 3}
\,
\left[ {V(T)\over \sqrt{1 -
\dot T^2}} + \Lambda \right] \, .
\ee
This is a constraint equation.
The time derivative of this equation is automatically satisfied once 
eqs.\refb{e8} is satisfied.

In order to solve these equations we need to know the initial conditions. 
Since there are two variables $T$ and $a$, satisfying second order 
differential 
equation \refb{e8}, we might naively expect that four initial conditions, 
giving the values of $T$, $\dot T$, $a$ and $\dot a$ at $x^0=0$, are 
needed 
to find 
the solution. However the Friedman equation \refb{e9} gives one 
constraint among the four initial conditions. This leaves us with three 
independent initial conditions. By choosing the origin of time suitably we 
can set either $T$ to $\dot T$ to zero at $x^0=0$, therby  
eliminating one more initial condition\cite{0203211,0203265}. We shall for 
definiteness consider the class of solutions where $\dot T=0$ at $x^0=0$. 
In this case we need two initial conditions to determine the solution 
completely. Without loss of generality we can take them to be the 
values of $T$ and $\dot a$ at $x^0=0$.

According to the open-closed duality conjecture, the decay of an unstable 
D-brane can be described completely by working in the open string field 
theory. In this theory, if we ignore the effect of massive open string 
modes, the inequivalent classical solutions describing 
the time evolution of a homogeneous tachyon field 
is described by one initial condition\cite{0203211,0203265}. For 
definiteness we shall take $\dot T=0$ at $x^0=0$ and use the 
value of the tachyon field $T$ at $x^0=0$ as the parameter labelling 
inequivalent solutions. Thus we see that the open string description of 
the rolling tachyon solution has one less parameter compared to the 
description in which we use both open and closed string fields ($T$ and 
$g_{\mu\nu}$) to describe the solution. This is not necessarily a 
contradiction, since as we have pointed out already, the open string 
description of the system does not have the capability of describing an 
arbitrary configuration in string theory, but can describe only a special 
class of 
configurations which describe the decay of a D-brane. What this 
mismatch of numbers indicates is that only a 
one parameter subspace of the two parameter family of solutions of 
eqs.\refb{e8}, \refb{e9} corresponds to the decay of a `pure D-brane' 
without any additional closed string background.
Fortunately in this case it is easy to identify this one parameter family 
of solutions. For this we note from eq.\refb{ess4} that the rolling 
tachyon 
solution 
describing the decay of a D-brane in open string theory is time reversal 
symmetric. Thus the solution of eqs.\refb{e8}, \refb{e9} representing the 
decay of a pure D-brane must also be time reversal 
symmetric.\footnote{Examples of time reversal symmetric cosmological 
solutions in the context of matrix model have been considered in 
\cite{0309138}.} This gives the initial conditions on $a$ and $T$ to be: 
\be \label{e10}
\dot a = 0, \quad \dot T=0, \quad T=T_0\, , \qquad \hbox{at} 
\quad x^0=0\, .
\ee
The Friedman equation now gives:
\be \label{e11}
k=1, \qquad a(x^0=0) = \left[ {8\pi G\over 3} \,
(V(T_0)+\Lambda)\right]^{-1/2}\equiv (H_0)^{-1}\, .
\ee
Since $G\sim g^2$ and $V(T_0)\sim g^{-1}$, we see that for small $g$ 
$H_0\sim g^{1/2}$.

The analysis of the equations of motion may be carried out as follows.  
Since initially $\dot T=0$, for sufficiently small $T_0$ there will be an 
interval of time during which $T$ remains fixed near $T_0$. The first 
equation in \refb{e8} and the initial condition \refb{e10} shows that 
during this period $a$ grows as
\be \label{e12}
a \simeq \cosh (H_0 x^0)\, ,
\ee
where $H_0$ has been defined in \refb{e11}.
In particular if $T_0=0$ then $T$ does not evolve, and \refb{e12} becomes 
exact. Since during this evolution the scale factor $a$ grows 
exponentially, the 
$k/a^2$ term in the Friedman equation \refb{e9} decays rapidly compared to 
the 
energy density in the tachyon field and the bulk cosmological constant 
$\Lambda$ as 
given by the right hand side of 
this equation.

For any non-zero $T_0$, however, the tachyon eventually starts evolving.
Once $T$ is of 
order 1, $V'/V$ is negative and of order 1. On the other hand, 
$\dot a /a$, which is of order $H_0$, is small in the weak coupling limit 
since $H_0$ is of order $g^{1/2}$.
Thus we expect the $\dot T \dot a/a$ term 
in the second equation in \refb{e8} to be small in magnitude compared to 
the $V'(T)/V(T)$ 
term. As a result the term inside the square bracket in the right hand 
side 
of this equation is 
negative and $\dot T=1$ is an attractive fixed point of this 
equation\cite{0209122,0303035}. 
As $T$ evolves and $\dot T$ approaches its limiting value 1, the energy 
density $V(T)/\sqrt{1-\dot T^2}$ associated with the tachyon field behaves 
like pressureless matter\cite{0204143,0209122} and falls off as $a^{-3}$. 
Since the 
curvature term $k/a^2$ in eq.\refb{e9}
falls off as $a^{-2}$, eventually the curvature term dominates over the 
tachyon matter term proportional to $V(T)/\sqrt{1-\dot T^2}$.

The crucial question that determines the subsequent evolution of the 
universe is: how does the cosmological constant term 
$\Lambda$ compare with the curvature term at this cross-over point? If the 
magnitude of the
curvature term is large compared to $\Lambda$ at this point, then soon 
after the cross-over, the curvature term overcomes the combined 
contribution from tachyon matter and $\Lambda$ and brings the expansion to 
a halt. After this the universe begins to recollapse, and eventually 
ends 
up in a big crunch. By time reversal symmetry such a universe must also 
have began with a big bang. On the other hand if the curvature term is 
small compared to the cosmological constant term at the cross-over point, 
then the right hand side of the Friedman equation never vanishes, 
and the universe continues to expand. Thus
there is no singularity in the future, and hence, by time reversal 
symmetry, no singularity in the past.

\begin{figure}[!ht]
\leavevmode
\begin{center}
\epsfysize=5cm
\epsfbox{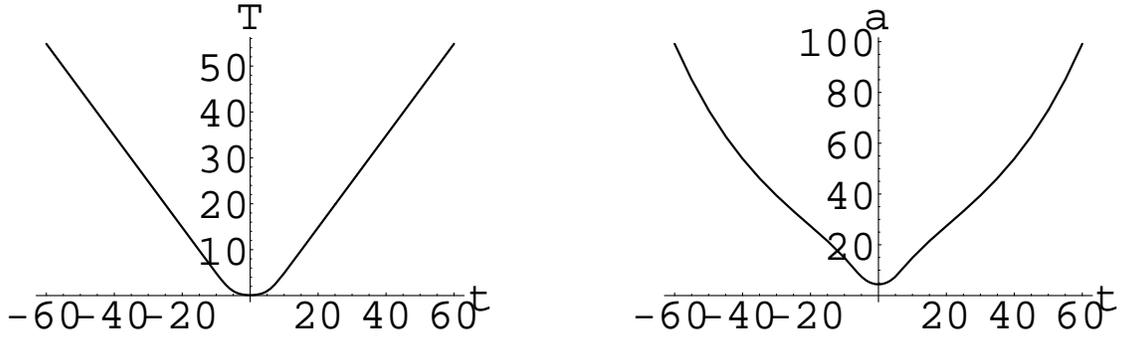}
\end{center}
\caption{Numerical results for the time evolution of $T$ and $a$ 
for $8\pi G V(0) / 3 = .05$, $8\pi G \Lambda / 3 = .001$,
$T_0=.1$. In this case $a$ is always positive and hence the solution 
is non-singular.} \label{f1}
\end{figure}

\begin{figure}[!ht]
\leavevmode
\begin{center}
\epsfysize=5cm
\epsfbox{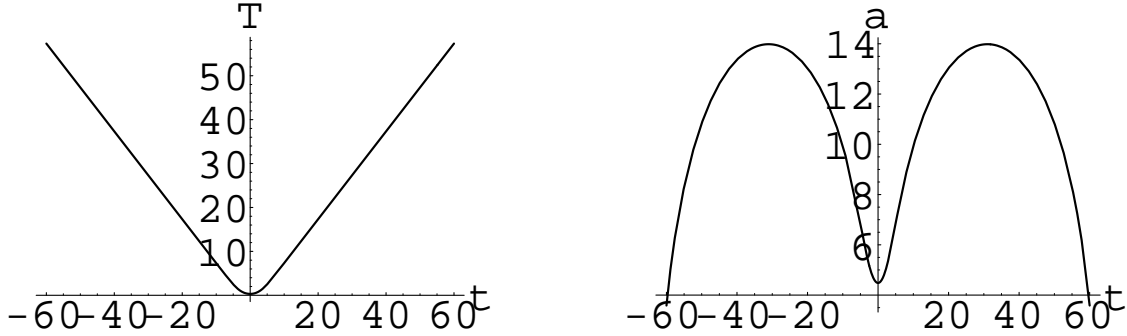}
\end{center}
\caption{Numerical results for the time evolution of $T$ and $a$
for $8\pi G V(0) / 3 = .05$, $8\pi G \Lambda / 3 = .001$,
$T_0=.32$. In this case $a$ 
vanishes at a finite time, making the solution
singular.} \label{f2}
\end{figure}

For any given $\Lambda$, which of these two scenarios takes place depends 
on the initial value $T_0$ of $T$. Suppose $a_f$ is the value of the 
scale factor at the end of the initial inflationary period. 
At this stage the energy density stored in the tachyon field is of the 
same order as the initial energy density $V_0\equiv V(0)$.
After this 
the tachyon contribution begins to behave like matter and its energy 
density falls 
off as 
$a^{-3}$. Then during further expansion we may estimate the tachyon 
contribution to the Friedman equation to be of order $V_0 a_f^3 / a^3$, 
and write the Friedman equation as:
\be \label{e13}
\left({\dot a \over a}\right)^2 \sim -{1\over a^2} + {8\pi G\over 3} 
\left[ 
{V_0 a_f^3\over 
a^3} 
+ \Lambda\right] \, .
\ee
The value of $a$ where the curvature term becomes comparable to the 
tachyon contribution is
\be \label{e14}
a_{cross} \sim G\, V_0 a_f^3\, .
\ee
Thus in order that at this point the $\Lambda$ term dominates over the 
curvature term, we need:
\be \label{e15}
\Lambda >> G^{-1} \, (a_{cross})^{-2} \sim G^{-3} (V_0)^{-2} 
(a_f)^{-6}\, .
\ee
$a_f$ is determined by the initial value $T_0$ of $T$. In particular 
smaller is the value of $T_0$, larger is the duration of the initial near 
exponential expansion, and larger is the value of $a_f$. Since 
$a_f=\infty$ for $T_0=0$, we see that 
for any $\Lambda$ there is a critical value $T_c$ such that for all 
$T_0<T_c$ we satisfy the inequality \refb{e15} and get a non-singular 
cosmological solution. In order to 
illustrate this point we have plotted in figs.~\ref{f1} and \ref{f2} the 
numerical results for the time evolution of $T$ and $a$ for 
$8\pi G V(0) / 3 = .05$, $8\pi G \Lambda / 3 = .001$, with two initial 
conditions on $T$: $T=.1$ at $x^0=0$ and $T=.32$ at $x^0=0$. For the first 
case $a$ is always positive and the solution is non-singular, whereas 
for the second case $a$ vanishes at some finite time and the solution hits 
a 
big-crunch singularity in the future. As is clear from these figures, due 
to time 
reversal symmetry of these solutions the
existence of the final singularity is correlated with the existence of the 
initial singularity. Thus a solution that is free from big crunch 
singularity is also free from big bang singularity, and describes a 
completely non-singular cosmology. On the other hand, a solution with a 
big crunch singularity also has a big bang singularity. Thus if we 
hypothesize that initial conditions in string theory avoid big 
bang singularity, then 
it automatically forces the universe to have sufficient inflation so as to 
avoid a big crunch singularity.

The above example illustrates how open closed duality conjecture can
restrict the class of allowed cosmological solutions describing the decay
of space-filling branes. We should emphasize again that this does not
imply that the solutions outside this class are not allowed, but just that
their description in the language of open string field theory is more
complicated and probably involves multiple D-branes.

Before we conclude we would like to add some cautionary remarks:
\begin{enumerate} 
\item We should keep in mind that the tachyon effective action
\refb{e7} is at best a qualitative description of the system, and even
then its range of validity is limited. In particular at late time it
always gives the equation of state $p/T_{00}=0$. However given the
interpretation of tachyon matter as a system of massive closed strings at
high density\cite{0305011,0306137}, we know that as the universe expands
and the density falls below the Hagedorn density, the equation of state of
the system should go from that of matter to that of radiation, and hence
the universe will expand as a radiation dominated universe. However the
same argument as given earlier shows that even in this case for 
sufficiently
small $T_0$ there will always be enough inflation at the beginning so that
by the time the curvature term becomes comparable to the radiation
contribution, the cosmological constant $\Lambda$ takes over and hence the
expansion continues.

\item The inflation produced by the 
solution is 
not a slow roll inflation, since $V''/V$ is of order unity.
Thus this solution cannot be used for a realistic model for 
inflationary cosmology. Nevertheless such solutions could have played a 
role in the early history of the universe, {\it e.g.} during a 
preinflationary phase.

\item We have not considered the possibility that the time reversal 
non-invariant classical solutions in tachyon effective field theory 
coupled to gravity, corresponding to putting arbitrary 
boundary condition on $\dot a$, could also correspond to solutions in open 
string field theory on a single space-filling D-brane, with complicated 
boundary conditions on the massive open string fields which do not respect 
time reversal symmetry. While this is certainly a possibilty, given 
our experience in $c=1$ matrix model that a 
generic closed string deformation cannot be described by an open string 
field configuration on a single D-brane, this seems unlikely.

\item Our analysis has been entirely classical. One could argue that when 
quantum corrections are taken into account then non-singular solutions of 
the type considered here, where the tachyon remains near the top of the 
potential for a long period, may not be possible due to quantum 
uncertainty. On the other hand, if we examine the quantum theory of an 
inverted harmonic oscillator (which approximates the tachyon dynamics in 
the open string theory near the top of the potential), the spectrum is 
continuous and there certainly is 
an eigenstate of the Hamiltonian with exactly zero eigenvalue. If 
open-closed duality conjecture holds, one could ask what will be the 
description of this state in the combined open-closed string theory. We 
hope that the wave-function of such a state will be peaked around the 
classical trajectory in the ($a$-$T$)-plane that we get by solving the 
coupled equations \refb{e8}-\refb{e11}. An analysis of the quantum system 
corresponding to the classical equations \refb{e8}-\refb{e11} might give 
us some insight into this issue.

\end{enumerate}

Finally we would like to note that in the case of the decay of unstable 
D-branes in $c=1$ string theory the correct computation of the closed 
string radiation from the brane 
is obtained by using the Hartle-Hawking 
prescription\cite{0303139,0304192}. We have arrived at the open-closed 
duality conjecture based on the results of this computation, and applied 
it to determine the initial condition for cosmological solutions 
describing the decay of an unstable brane. It will be interesting to 
explore any possible relation that might exist between this proposal and 
the Hartle-Hawking prescription\cite{HH} for determining the allowed 
wave-functions
of the universe.

\medskip

{\bf Acknowledgement:} I would like to thank the organisers and 
participants of the Nobel 
Symposium on Cosmology and String Theory and IIT Kanpur workshop on String 
Theory for a stimulating environment. I 
would also like to thank Arvind Borde for useful communications.

\end{document}